\documentclass[aps,prb,twocolumn,showpacs,amsmath,amssymb]{revtex4}
\usepackage{graphicx}
\usepackage{bm}

\def\br{ \bm{r} }

\def\bk{ \bm{k} }
\def\bK{ \bm{K} }
\def\bp{ \bm{p} }

\def\bq{ \bm{q} }

\def\bB{ \bm{B} }
\def\bv{ \bm{v} }
\def\bgam{ \bm{\gamma} }

\def\im{ \,\mathrm{Im}\, }

\def\sign{ \,\mathrm{sign}\, }

\begin{document}

\title{Effects of impurities on superconductivity in noncentrosymmetric compounds}

\author{V. P. Mineev$^{1}$ and K. V. Samokhin$^{1,2}$}

\affiliation{$^{1}$ Commissariat \`a l'Energie Atomique,
DSM/DRFMC/SPSMS, 38054 Grenoble,
France\\
$^{2}$ Department of Physics, Brock University, St.Catharines,
Ontario L2S 3A1, Canada}
\date{\today}

\begin{abstract}
We microscopically derive the Ginzburg-Landau free energy
functional for a noncentrosymmetric superconductor with a large
spin-orbit splitting of the electron bands, in the presence of
nonmagnetic impurities. The critical temperature is found to be
suppressed by disorder, both for conventional and unconventional
pairing, in the latter case according to the universal
Abrikosov-Gor'kov function. The impurity effect on the upper
critical field turns out to be non-universal, determined by the
pairing symmetry and the band structure. In a BCS-like model,
$T_c$ is not affected, while $H_{c2}$ increases with disorder. For
unconventional pairing, both $T_c$ and $H_{c2}$ are suppressed by
disorder.
\end{abstract}

\pacs{74.20.-z, 74.25.Ha, 74.62.Dh}

\maketitle

\section{Introduction}
\label{sec: Intro}

The recent discovery of superconductivity in a heavy-fermion
compound CePt$_3$Si, see Ref. \onlinecite{Bauer04}, has renewed
interest, both experimental and theoretical, in the properties of
superconductors without inversion symmetry. The list of such
materials has been steadily growing in recent months and now also
includes UIr (Ref. \onlinecite{Akazawa04}), CeRhSi$_3$ (Ref.
\onlinecite{Kimura05}), CeIrSi$_3$ (Ref. \onlinecite{Sugitani06}),
Y$_2$C$_3$ (Ref. \onlinecite{Amano04}), and
Li$_2$(Pd$_{1-x}$,Pt$_x$)$_3$B (Ref. \onlinecite{LiPt-PdB}).

A peculiar property of noncentrosymmetric crystals is that the
spin-orbit (SO) coupling qualitatively changes the nature of
single-electron states, namely it leads to the lifting of spin
degeneracy and the splitting of the energy bands. This has
important consequences for superconductivity. If the typical band
splitting $E_{SO}$ is smaller than the superconducting critical
temperature, then the effects of the SO coupling can be treated
perturbatively. In particular, the pairing interaction can be
chosen to be a function of the quasiparticle spins and momenta
near the Fermi surface unaffected by the SO coupling. This
approach to the theory of noncentrosymmetric superconductivity was
introduced by Edelstein in Ref. \onlinecite{Edel89} and further
developed in Refs. \onlinecite{Edel95,Edel96,FAKS04,FAMS05}. Due
to the absence of inversion symmetry the superconducting order
parameter does not, in general, have definite parity, becoming a
mixture of spin-singlet and spin-triplet components. The triplet
component appears, however, only when a spin-triplet channel is
explicitly present in the pairing interaction, even for a finite
value of the SO band splitting. When the SO coupling increases,
only a certain type of the triplet pairing survives, along with
the singlet component.

In the limit of strong SO coupling, i.e. when the band splitting
exceeds the superconducting energy scales, the Cooper pairing
between the electrons with opposite momenta occurs only if they
are from the same nondegenerate band. Interband pairing is still
possible, but only at the band degeneracy lines or points in the
momentum space, and can be neglected. According to Ref.
\onlinecite{SZB04}, this is the scenario that is realized in
CePt$_3$Si: The band structure calculations show that $E_{SO}$
ranges from 500K to 2000K, which is much larger than the critical
temperature $T_c=0.75$K, even taking into account a possible mass
renormalization due to the strong correlation effects. The same is
likely to be the case in other materials, e.g. Li$_2$Pd$_3$B and
Li$_2$Pt$_3$B, see Ref. \onlinecite{LP05}. The pairing interaction
in the strong SO coupling case is most naturally introduced using
the basis of the exact band states,\cite{SZB04,GR01,SC04,Min04}
which already incorporate all the effects of the crystal lattice
potential and the SO coupling.

In the band representation, the superconducting order parameter is
given by a set of complex gap functions, one for each band, which
are coupled due to the interband scattering of the Cooper pairs
and other mechanisms, e.g. impurity scattering. The overall
structure of the theory resembles that of multi-band
superconductors.\cite{SMW59,Moskal59} However, since the bands are
nondegenerate, the pairing symmetry is peculiar: While each order
parameter is an odd function of momentum, the gap symmetry, in
particular the positions of the nodes, is determined by one of the
even representations of the point group of the crystal. When
expressed in the spin representation, the order parameter becomes
a mixture of singlet and triplet components,\cite{Min04} the
latter appearing even without any spin-triplet term in the pairing
interaction, as an inevitable consequence of the SO band splitting
and the difference between the gap magnitudes and the densities of
states in different bands.

Noncentrosymmetric superconductors exhibit a variety of unusual
features, which are absent in the centrosymmetric case, such as a
strongly anisotropic spin susceptibility with a large residual
component,\cite{Bul76,Edel89,GR01,Yip02,FAKS04,FAS04,Sam05,Min05}
magnetoelectric
effect,\cite{Lev85,Edel89,Edel95,Edel96,Yip02,Fuji05,Edel05} and
helical superconducting
states.\cite{Min93,MinSam94,Agter03,DF03,Sam04,Kaur05,Oka06} Other
properties of interest include the nuclear spin
relaxation,\cite{SamNMR,Hayashi06-2} Josephson and quasiparticle
tunnelling,\cite{Yokoyama05,Borkje06} electron correlation
effects,\cite{Fuji05,Fuji06} superfluid density and the London
penetration depth,\cite{Hayashi06} and the free energy in the
clean case.\cite{Sam04,Edel96}

In this article we study the effects of nonmagnetic impurities on
the superconducting properties. In contrast to the treatment of
the same problem in Refs. \onlinecite{FAMS05} and
\onlinecite{Edel05}, we assume that the SO coupling is larger than
the superconducting energy scales, which necessitates using the
band representation of the pairing Hamiltonian. Our main goal is
to find how the superconducting critical temperature and the upper
critical field depend on the impurity concentration, for both
conventional and unconventional pairing. The article is organized
as follows. In Sec. \ref{sec: normal} we obtain the band
representations of the impurity Hamiltonian and of the
disorder-averaged Green's function of electrons in the normal
state. In Sec. \ref{sec: SC}, we derive the Ginzburg-Landau (GL)
free energy functional in the superconducting state. In Sec.
\ref{sec: observables}, we calculate the critical temperature and
the upper critical field in the presence of impurities and also
discuss the application of our results to a model of
superconductivity in CePt$_3$Si.

\section{Impurity scattering in normal state}
\label{sec: normal}

In a noncentrosymmetric crystal with SO coupling the electron
bands are nondegenerate. The formal reason is that without the
inversion operation one cannot, in general, have two orthogonal
degenerate Bloch states at the same wave vector $\bk$. In the
limit of zero SO coupling there is an additional symmetry in the
system -- the invariance with respect to arbitrary rotations in
spin space -- which preserves two-fold degeneracy of the bands.
Let us consider one spin-degenerate band with the dispersion given
by $\epsilon_0(\bk)$, and turn on the SO coupling. The Hamiltonian
of noninteracting electrons in the presence of scalar impurities
can be written in the form $H=H_0+H_{imp}$, where
\begin{equation}
\label{H_SO}
    H_0=\sum\limits_{\bk,\alpha\beta}[\epsilon_0(\bk)\delta_{\alpha\beta}+
    \bgam(\bk)\bm{\sigma}_{\alpha\beta}]
    a^\dagger_{\bk\alpha}a_{\bk\beta},
\end{equation}
$\alpha,\beta=\uparrow,\downarrow$ denote the spin projection on
the $z$-axis, $\sum_{\bk}$ stands for the summation over the first
Brillouin zone, $\hat{\bm{\sigma}}$ are the Pauli matrices, the
chemical potential is included in $\epsilon_0(\bk)$, and
\begin{equation}
\label{H_imp}
    H_{imp}=\int d^3\br\sum_\alpha
    U(\br)\psi^\dagger_\alpha(\br)\psi_\alpha(\br).
\end{equation}
The impurity  potential $U(\br)$ is a random function with zero
mean and the correlator $\langle
U(\br_1)U(\br_2)\rangle=n_{imp}U_0^2\delta(\br_1-\br_2)$, where
$n_{imp}$ is the impurity concentration, and $U_0$ is the strength
of an individual point-like impurity.

The ``bare'' band dispersion satisfies
$\epsilon_0(-\bk)=\epsilon_0(\bk)$,
$\epsilon_0(g^{-1}\bk)=\epsilon_0(\bk)$, where $g$ is any
operation from the point group $\mathbb{G}$ of the crystal. The SO
coupling of electrons with the crystal lattice is described by the
pseudovector $\bgam(\bk)$, which has the following symmetry
properties: $\bgam(\bk)=-\bgam(-\bk)$,
$(g\bgam)(g^{-1}\bk)=\bgam(\bk)$. For example, the point symmetry
of CePt$_3$Si, CeRhSi$_3$ and CeIrSi$_3$ is tetragonal and
described by $\mathbb{G}=\mathbf{C}_{4v}$, which is generated by
the rotations $C_{4z}$ about the $z$ axis by an angle $\pi/2$ and
the reflections $\sigma_x$ in the vertical plane $(100)$. The
pseudovector $\bgam(\bk)$ can be written as
\begin{equation}
\label{gamma_C4v}
    \bgam(\bk)=\gamma_\perp[\bm{\phi}_{E,u}(\bk)\times\hat z]
    +\gamma_\parallel\phi_{A_2,u}(\bk)\hat z,
\end{equation}
where $\gamma_\perp$ and $\gamma_\parallel$ are constants, and
$\bm{\phi}_{E,u}$ and $\phi_{A_2,u}$ are the odd basis functions
of the irreducible representations $E$ (two-dimensional) and $A_2$
(one-dimensional) of $\mathbf{C}_{4v}$.\cite{Sam04} The
Hamiltonian (\ref{H_SO}) with $\bgam(\bk)$ given by Eq.
(\ref{gamma_C4v}) is a three-dimensional generalization of the
Rashba model, which is widely used to describe the effects of SO
coupling in two-dimensional semiconductor
heterostructures.\cite{Rashba60}

The SO coupling strength depends on the quasiparticle momentum and
might vanish, for symmetry reasons, along some directions or at
some isolated points in the Brillouin zone. The former possibility
is realized in the tetragonal case: $\bgam(\bk)=0$ along the axis
$k_x=k_y=0$, which can be seen from the polynomial expressions for
the basis functions: $\bm{\phi}_{E,u}(\bk)\sim(k_x,k_y)$ and
$\phi_{A_2,u}(\bk)\sim k_xk_yk_z(k_x^2-k_y^2)$. However, this is
not generic: For example, the point symmetry of
Li$_2$(Pd$_{1-x}$,Pt$_x$)$_3$B is described by the cubic group
$\mathbb{G}=\mathbf{O}$, which contains only the rotations about
the axes of the second, third, and fourth order. In this case the
function $\bgam(\bk)$ transforms according to the vector
representation $F_1$:
\begin{equation}
\label{gamma_O}
    \bgam(\bk)=\gamma_0\bm{\phi}_{F_1,u}(\bk).
\end{equation}
The representative expression for the basis function is simply
$\bm{\phi}_{F_1,u}(\bk)\sim(k_x,k_y,k_z)$, so that the SO coupling
vanishes at the point $k_x=k_y=k_z=0$.

The diagonalization of $H_0$ yields the following eigenvalues and
eigenstates:
\begin{equation}
\label{Rashba_bands}
    \xi_\lambda(\bk)=\epsilon_0(\bk)+\lambda|\bgam(\bk)|,
\end{equation}
and
\begin{equation}
\label{chis}
    \chi_{\bk\lambda}(\br,\alpha)=\frac{1}{\sqrt{{\cal V}}}u_{\alpha\lambda}(\bk)e^{i\bk\br},
\end{equation}
where $\lambda=\pm$ is the band index, ${\cal V}$ is the system
volume,
\begin{equation}
\label{Rashba_spinors}
    \begin{array}{l}
    \displaystyle u_{\uparrow\lambda}(\bk)=e^{i\theta_\lambda}
    \sqrt{\frac{|\bgam|+\lambda\gamma_z}{2|\bgam|}},\\
    \displaystyle u_{\downarrow\lambda}(\bk)=\lambda e^{i\theta_\lambda}
    \frac{\gamma_x+i\gamma_y}{\sqrt{2|\bgam|(|\bgam|+\lambda\gamma_z)}},
    \end{array}
\end{equation}
and $\theta_\lambda$ are arbitrary (in general $\bk$-dependent)
phases. The Bloch spinor components $u_{\alpha\lambda}$ form a
unitary matrix $\hat u(\bk)$. It follows from Eq.
(\ref{Rashba_bands}) that in the presence of SO coupling the
degeneracy of the electron bands is lifted everywhere in the
Brillouin zone, except maybe for some high-symmetry lines or
points, where $\bgam(\bk)=0$ (the band structure might be such
that the zeros of $\bgam$ are not located on the Fermi surface).
For the typical band splitting $E_{SO}$ one can use, for instance,
the Fermi-surface average of $2|\bgam(\bk)|$.

The band representation of the free-electron Hamiltonian has the
following form:
\begin{equation}
\label{H_0_band}
    H_0=\sum_{\bk}\sum_{\lambda=\pm}\xi_\lambda(\bk)c^\dagger_{\bk\lambda}c_{\bk\lambda}.
\end{equation}
The band dispersion functions (\ref{Rashba_bands}) are even in
$\bk$ due to time reversal symmetry: the states
$\chi_{\bk\lambda}$ and $K\chi_{\bk\lambda}$ belong to $\bk$ and
$-\bk$, respectively, and have the same energy. Here
$K=i\hat\sigma_2K_0$ is the time reversal operation, and $K_0$ is
the complex conjugation.

Writing the field operators in Eq. (\ref{H_imp}) in the form
$\psi_\alpha(\br)=\sum_{\bk,\lambda}\chi_{\bk\lambda}(\br,\alpha)c_{\bk\lambda}$,
we obtain the band representation of the impurity Hamiltonian:
\begin{equation}
\label{H_imp_band}
    H_{imp}=\frac{1}{{\cal V}}\sum_{\bk\bk'}\sum_{\lambda\lambda'}
    \tilde U_{\lambda\lambda'}(\bk,\bk')
    c^\dagger_{\bk\lambda}c_{\bk'\lambda'},
\end{equation}
where
\begin{equation}
\label{tilde U k}
    \tilde U_{\lambda\lambda'}(\bk,\bk')=U(\bk-\bk')
    w_{\lambda\lambda'}(\bk,\bk'),
\end{equation}
$U(\bq)$ is the Fourier transform of the impurity potential,
$\langle U(\bq)U(\bq')\rangle=n_{imp}U_0^2{\cal
V}\delta_{\bq,-\bq'}$, and
\begin{equation}
\label{w}
    \hat w(\bk,\bk')=\hat u^\dagger(\bk)\hat u(\bk').
\end{equation}
We see that the impurity scattering amplitude acquires both
intraband and interband contributions and also becomes
anisotropic, even for isotropic scalar impurities. In more general
cases, for instance in the presence of spin-dependent scattering
in Eq. (\ref{H_imp}), the band representation of the impurity
Hamiltonian still has the form (\ref{H_imp_band}), but with a
different $\tilde U_{\lambda\lambda'}(\bk,\bk')$.

We note that the matrix elements of the anisotropy factor $\hat w$
satisfy the following useful identity:
\begin{equation}
\label{w identity}
    |w_{\lambda\lambda'}(\bk,\bk')|^2=
    \frac{1+\lambda\lambda'\hat\bgam(\bk)\hat\bgam(\bk')}{2},
\end{equation}
which can be easily verified using the explicit expressions
(\ref{Rashba_spinors}) for the Bloch spinors.

The electron Green's function in the band representation is
introduced in the standard fashion:\cite{AGD}
\begin{equation}
\label{G def}
    G_{\lambda\lambda'}(\bk,\tau;\bk',\tau')=-\langle T_\tau
    c_{\bk\lambda}(\tau)c^\dagger_{\bk'\lambda'}(\tau')\rangle.
\end{equation}
In the absence of impurities it has the following form:
\begin{equation}
\label{G_0}
    G_{0,\lambda\lambda'}(\bk,\omega_n)=\frac{\delta_{\lambda\lambda'}}{i\omega_n-\xi_\lambda(\bk)},
\end{equation}
where $\omega_n=(2n+1)\pi T$ is the fermionic Matsubara frequency
(we use the units in which $k_B=1$). Now we will show that, even
when the disorder is included, the average Green's function
remains band-diagonal.

\begin{figure}
    \includegraphics[width=5cm]{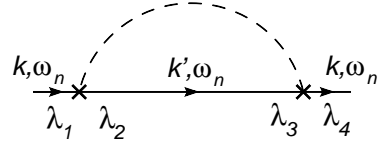}
    \caption{The impurity self-energy in the band representation. The dashed line
    corresponds to $n_{imp}U_0^2$, the vertices include the anisotropy factors $\hat w(\bk,\bk')$,
    and the solid line is the average Green's function of electrons in the normal state.
    It is shown in the
    text that the self-energy is nonzero only if $\lambda_1=\lambda_4$ and $\lambda_2=\lambda_3$.}
    \label{fig: self-energy}
\end{figure}

The disorder averaging with the Hamiltonian (\ref{H_imp_band}) can
be performed using the standard methods, see Ref.
\onlinecite{AGD}. The result is the matrix Dyson equation of the
form $\hat G^{-1}=\hat G_0^{-1}-\hat\Sigma$, where $\hat G$ is the
average Green's function and $\hat\Sigma$ is the impurity
self-energy. In the Born approximation,
\begin{eqnarray}
\label{Sigma_n_Born}
    &&\hat\Sigma(\bk,\omega_n)=n_{imp}U_0^2\nonumber\\
    &&\qquad\times\int\frac{d^3\bk'}{(2\pi)^3}
    \hat w(\bk,\bk')\hat G(\bk',\omega_n)\hat w(\bk',\bk)
\end{eqnarray}
(here we have taken the thermodynamic limit ${\cal V}\to\infty$).
The diagrammatic representation of the self-energy is given in
Fig. \ref{fig: self-energy}. We seek solution of the Dyson
equation in a band-diagonal form:
$G_{\lambda\lambda'}=G_\lambda\delta_{\lambda\lambda'}$, then the
integrand on the right-hand side of Eq. (\ref{Sigma_n_Born})
becomes
\begin{eqnarray}
\label{UGU}
    &&\hat u(\bk')\hat G(\bk',\omega_n)\hat u^\dagger(\bk')\nonumber\\
    &&\qquad=
    \frac{G_+(\bk',\omega_n)+G_-(\bk',\omega_n)}{2}\hat\tau_0\nonumber\\
    &&\qquad+\frac{G_+(\bk',\omega_n)-G_-(\bk',\omega_n)}{2}\hat\zeta(\bk'),
\end{eqnarray}
where $\hat\tau_i$ are the Pauli matrices, and
$\hat\zeta(\bk)=\hat u(\bk)\hat\tau_3\hat u^\dagger(\bk)$. Using
the expressions (\ref{Rashba_spinors}), one obtains
$\hat\zeta(\bk)=\hat\bgam(\bk)\hat{\bm{\tau}}$, which is an odd
function of $\bk$ ($\hat\bgam=\bgam/|\bgam|$). Therefore the last
line in Eq. (\ref{UGU}) vanishes after the momentum integration,
and $\hat\Sigma(\bk,\omega_n)=\Sigma(\omega_n)\hat\tau_0$. The
real part of the self-energy renormalizes the chemical potential,
and for the imaginary part we obtain:
$$
    \im\Sigma(\omega_n)=-\pi n_{imp}U_0^2\frac{N_++N_-}{2}\sign\omega_n,
$$
where $N_\lambda={\cal V}^{-1}\sum_{\bk}\delta[\xi_\lambda(\bk)]$
is the Fermi-level density of states in the $\lambda$th band. In
order to preserve the usual form of the impurity self-energy in
the normal state, we introduce the notation $N_F=(N_++N_-)/2$ and
also define the elastic mean free time as $\tau=(2\pi
n_{imp}U_0^2N_F)^{-1}$. Then the average Green's function takes
the form
\begin{equation}
\label{G_n}
    G_{\lambda\lambda'}(\bk,\omega_n)=
    \frac{\delta_{\lambda\lambda'}}{i\omega_n-\xi_\lambda(\bk)+(i/2\tau)\sign\omega_n}.
\end{equation}

The derivation above is valid under the assumption that the
elastic scattering rate is small compared with the Fermi energy
$\epsilon_F$, which justifies neglecting the diagrams with crossed
impurity lines in the self-energy in Fig. \ref{fig: self-energy}.

\subsection{Effects of magnetic field}
\label{sec: B in normal state}

In the presence of a nonzero uniform magnetic field $\bB$ Eq.
(\ref{H_0_band}) is replaced by
\begin{equation}
\label{H0 B}
    H_0=\sum\limits_{\bk,\lambda}c^\dagger_{\bk\lambda}
    {\cal E}_{\lambda}(\bk)c_{\bk\lambda},
\end{equation}
where ${\cal E}_\lambda$ is the effective band Hamiltonian in the
$\bk$-space.\cite{LL9} It was first pointed out by
Peierls,\cite{Pei33} based on the requirement of gauge invariance,
that the magnetic field can be included in the band electron
theory by simply replacing the wave vector $\bk$ in the zero-field
band dispersion $\xi_\lambda(\bk)$ by the operator
$\bK=\bk+(e/\hbar c)\bm{A}(\hat\br)$, where
$\hat\br=i\bm{\nabla}_{\bk}$ is the position operator in the
$\bk$-representation and $e$ is the absolute value of the electron
charge. We use the symmetric gauge, in which
$$
    \bK=\bk+i\frac{e}{2\hbar c}(\bB\times\bm{\nabla}_{\bk}).
$$
The Peierls substitution gives only the zero-order term in the
expansion of the effective single-band Hamiltonian in powers of
$\bB$, which in our case can be written as
\begin{equation}
\label{H_eff}
    {\cal E}_\lambda(\bk)=\xi_\lambda(\bK)-
    \bB\bm{m}_\lambda(\bK)+....
\end{equation}
The second term on the right-hand side is the analog of the Zeeman
interaction for nondegenerate bands.\cite{Sam05} In the
generalized Rashba model (\ref{H_SO}) it has the following form:
\begin{equation}
\label{mu_Rashba}
    \bm{m}_\lambda(\bk)=\lambda\mu_B\hat\bgam(\bk)
\end{equation}
($\mu_B$ is the Bohr magneton), which is valid everywhere except
for the vicinity of the band crossing points, where the
approximation of independent nondegenerate bands fails and the
effective band Hamiltonian approach might not work.

It is convenient to introduce the Fourier transform of the Green's
function (\ref{G def}):
\begin{equation}
\label{G rr def}
    G_{\lambda\lambda'}(\br,\br';\omega_n)=\frac{1}{{\cal V}}\sum\limits_{\bk\bk'}
    e^{i\bk\br-i\bk'\br'}
    G_{\lambda\lambda'}(\bk,\bk';\omega_n).
\end{equation}
We would like to stress that this is not the same as the electron
Green's function in the coordinate-spin representation. The latter
is defined as
$\langle\br\alpha|(i\omega_n-H)^{-1}|\br'\beta\rangle=
\sum_{\bk\bk',\lambda\lambda'}\langle\br\alpha|\bk\lambda\rangle
G_{\lambda\lambda'}(\bk,\bk';\omega_n)\langle\bk'\lambda'|\br'\beta\rangle$,
where
$\langle\br\alpha|\bk\lambda\rangle=\chi_{\bk\lambda}(\br,\alpha)$
is the spinor wave function (\ref{chis}), with
$\alpha,\beta=\uparrow,\downarrow$.

The Green's function (\ref{G rr def}) satisfies the equation
\begin{equation}
\label{GF equation r}
    (i\omega_n\hat\tau_0-\hat h_0-\hat h_{imp})\hat
    G(\br,\br';\omega_n)=\hat\tau_0\delta(\br-\br').
\end{equation}
Here $h_{0,\lambda\lambda'}=\delta_{\lambda\lambda'}{\cal
E}_\lambda(\bK_{\br})$ is obtained by replacing $\bK$ in Eq.
(\ref{H_eff}) by
\begin{equation}
\label{hat K}
    \bK_{\br}=-i\bm{\nabla}_{\br}+\frac{e}{\hbar
    c}\bm{A}(\br)=-i\bm{\nabla}_{\br}+\frac{e}{2\hbar
    c}(\bB\times\br),
\end{equation}
and $\hat h_{imp}$ is the Fourier transform of the impurity
Hamiltonian (\ref{H_imp_band}). The latter is nonlocal in real
space and can be represented as a differential operator of
infinite order:
\begin{eqnarray}
\label{h_imp G}
    &&(\hat h_{imp}\hat G)_{\lambda\lambda'}(\br,\br';\omega_n)=\sum_{\lambda_1}
    [{\cal U}_{0,\lambda\lambda_1}(\br)\nonumber\\
    &&\qquad+{\bm{{\cal U}}}_{1,\lambda\lambda_1}(\br)\bK_{\br}+...]
    G_{\lambda_1\lambda'}(\br,\br';\omega_n),
\end{eqnarray}
where
\begin{eqnarray*}
    {\cal U}_{0,\lambda\lambda'}(\br)&=&\frac{1}{{\cal V}}\sum_{\bp}e^{i\bp\br}
    \tilde U_{\lambda\lambda'}\left(\frac{\bp}{2},-\frac{\bp}{2}\right),\\
    \bm{{\cal U}}_{1,\lambda\lambda'}(\br)&=&\frac{1}{{\cal V}}\sum_{\bp}e^{i\bp\br}\\
    &&\times 2(\bm{\nabla}_{\bp}-\bm{\nabla}_{\bp'})
    \left.\tilde
    U_{\lambda\lambda'}\left(\frac{\bp}{2},-\frac{\bp'}{2}\right)\right|_{\bp'=\bp},
\end{eqnarray*}
\emph{etc}. The magnetic field affects the electron Green's
function in Eq. (\ref{GF equation r}) through the vector potential
in the operators $\bK_{\br}$ (the ``orbital effect'') as well as
through the $\bm{m}$-term in Eq. (\ref{H_eff}) (the ``paramagnetic
effect'').

In the absence of impurities the solution of Eq. (\ref{GF equation
r}) is band-diagonal and can be represented in a factorized
form:\cite{Gor59}
\begin{eqnarray}
\label{GF factorized}
    G_{\lambda\lambda'}(\br,\br';\omega_n)&=&
    \bar G_{\lambda\lambda'}(\br,\br',\omega_n)\nonumber\\
    &&\times\exp\left[i\frac{e}{\hbar c}\int_{\br}^{\br'}\bm{A}(\br)d\br\right],
\end{eqnarray}
where the integration in the phase factor is performed along a
straight line connecting $\br$ and $\br'$, and $\bar
G_{\lambda\lambda'}(\br,\br',\omega_n)=\delta_{\lambda\lambda'}\bar
G_{0,\lambda}(\br-\br',\omega_n)$. In superconductors the orbital
effect of magnetic field on $\bar G_0$ can usually be
neglected,\cite{neglect orbital} and one obtains:
\begin{equation}
\label{bar G0}
    \bar G_{0,\lambda}(\bk,\omega_n)=
    \frac{1}{i\omega_n-\xi_\lambda(\bk)+\bm{m}_\lambda(\bk)\bB},
\end{equation}
see also Ref. \onlinecite{Sam04}.

The same argument can be used also in the disordered case, to show
that the Green's function before disorder averaging has the form
(\ref{GF factorized}), where $\bar G_{\lambda\lambda'}$ satisfies
Eq. (\ref{GF equation r}), in which the vector potential is
formally set to zero, but the paramagnetic term is still present.
In order to perform disorder averaging of $\bar
G_{\lambda\lambda'}$, we go back into the band-momentum
representation (\ref{G def}) and repeat the steps leading to Eq.
(\ref{Sigma_n_Born}). The only difference is that now the Green's
function is no longer even in $\bk$ due to the presence of the
paramagnetic term with  $\bm{m}_\lambda(\bk)=-
\bm{m}_\lambda(-\bk)$, see Eq. (\ref{bar G0}). Therefore the last
line in Eq. (\ref{UGU}) does not vanish identically after the
momentum integration, and the impurity self-energy is no longer
band-diagonal. One can show however that the off-diagonal
contributions to $\Sigma_{\lambda\lambda'}$ are of the order of
$\mu_BBN_F'$, which can be neglected. Therefore we arrive at the
following expression for the impurity-averaged $\bar
G_{\lambda\lambda'}$:
\begin{equation}
\label{bar G average gen}
    \langle\bar
    G_{\lambda\lambda'}(\bk,\bk';\omega_n)\rangle_{imp}=\delta_{\lambda\lambda'}
    \delta_{\bk,\bk'}\bar G_{\lambda}(\bk,\omega_n),
\end{equation}
where
\begin{eqnarray}
\label{bar G average}
    &&\bar G_{\lambda}(\bk,\omega_n)\nonumber\\
    &&\quad=\frac{1}{i\omega_n-\xi_\lambda(\bk)
    +\bm{m}_\lambda(\bk)\bB+(i/2\tau)\sign\omega_n}.\quad
\end{eqnarray}
While the orbital effect of the field is reduced to the phase
factor in the Green's function, see Eq. (\ref{GF factorized}), the
Zeeman interaction survives in $\bar G_{\lambda\lambda'}$ and
affects the quasiparticle energies. The Zeeman term will be
eventually dropped, because its effect on the GL free energy turns
out to be negligible, see the next section.

\section{Superconducting state}
\label{sec: SC}

Now let us take into account an attractive interaction between
electrons in the Cooper channel, using the basis of the exact
eigenstates of the noninteracting problem. The total Hamiltonian
is given by $H=H_0+H_{imp}+H_{int}$, where the first two terms are
given by Eqs. (\ref{H_0_band}) and (\ref{H_imp_band}), and the
last term has the following form:
\begin{eqnarray}
\label{H int}
    H_{int}=\frac{1}{2{\cal V}}\sum\limits_{\bk\bk'\bq}\sum_{\lambda\lambda'}
    V_{\lambda\lambda'}(\bk,\bk')c^\dagger_{\bk+\bq/2,\lambda}
    c^\dagger_{-\bk+\bq/2,\lambda}\nonumber\\
    \times c_{-\bk'+\bq/2,\lambda'}c_{\bk'+\bq/2,\lambda'}.
\end{eqnarray}
We assume, in the spirit of the Bardeen-Cooper-Schrieffer (BCS)
theory, that the pairing interaction is nonzero only inside the
thin shells of width $\varepsilon_c$ in the vicinity of the Fermi
surfaces, i.e. when
$|\xi_{\lambda}(\bk)|,|\xi_{\lambda'}(\bk')|\leq\varepsilon_c$.
The pairing potential satisfies
$V_{\lambda\lambda'}(-\bk,\bk')=V_{\lambda\lambda'}(\bk,-\bk')=-V_{\lambda\lambda'}(\bk,\bk')$,
which follows from the anti-commutation of the fermionic
operators. The diagonal elements of the matrix $\hat V$ describe
the intraband Cooper pairing, while the off-diagonal ones
correspond to the pair scattering from one band to the other. The
SO splitting of the bands is assumed to be large compared with all
the energy scales associated with superconductivity. In this case
the formation of the pairs of electrons belonging to different
bands is strongly suppressed. Although the bands may touch at some
isolated points at the Fermi surface, the interband pairing in the
vicinity of those points is still suppressed due to the phase
space limitations.

The pairing potential can be represented in a factorized form:
\begin{equation}
\label{pairing potential}
    V_{\lambda\lambda'}(\bk,\bk')=-V_{\lambda\lambda'}
    t_\lambda(\bk)t^*_{\lambda'}(\bk')\phi_\lambda(\bk)
    \phi^*_{\lambda'}(\bk'),
\end{equation}
where the coupling constants $V_{\lambda\lambda'}$ form a
symmetric positive-definite $2\times 2$ matrix,
$t_\lambda(\bk)=-t_\lambda(-\bk)$ are non-trivial phase factors,
see Refs. \onlinecite{GR01,SC04}, and
$\phi_\lambda(\bk)\equiv\phi^{(\lambda)}_{\Gamma,g}(\bk)$ are even
basis functions of an irreducible representation $\Gamma$ of the
point group $\mathbb{G}$ of the crystal (we keep only the
representation which corresponds to the pairing channel with the
maximum critical temperature).\cite{Book} While $\phi_+(\bk)$ and
$\phi_-(\bk)$ have the same symmetry, their momentum dependence
does not have to be exactly the same. We consider only
one-dimensional representations of $\mathbb{G}$, because, to the
best of our knowledge, there is no experimental evidence of
multi-dimensional superconductivity in real noncentrosymmetric
materials.

The basis functions are nonzero only inside the BCS shells and are
normalized: $\langle|\phi_\lambda(\bk)|^2\rangle_\lambda=1$, where
the angular brackets denote the averaging over the Fermi surface
in the $\lambda$th band:
$$
    \langle(...)\rangle_\lambda=\frac{1}{N_\lambda}
    \frac{1}{{\cal V}}\sum_{\bk}(...)\delta[\xi_\lambda(\bk)].
$$
The phase factors $t_\lambda(\bk)$ originate in the expression for
the time reversal operation for nondegenerate band electrons:
\begin{equation}
\label{t_def}
    K|\bk\lambda\rangle=t_\lambda(\bk)|-\bk,\lambda\rangle.
\end{equation}
For instance, for the eigenstates (\ref{Rashba_spinors}) we
obtain:
\begin{eqnarray}
    t_\lambda(\bk)=\lambda
    e^{-i[\theta_\lambda(\bk)+\theta_\lambda(-\bk)]}
    \frac{\gamma_x(\bk)-i\gamma_y(\bk)}{\sqrt{\gamma_x^2(\bk)+\gamma_y^2(\bk)}}.
\end{eqnarray}
Since the factors $t_\lambda$ explicitly depend on the arbitrary
phases $\theta_\lambda(\bk)$, they must drop out of the final
expressions for all observable quantities.

It is instructive to see how the phase factors $t_\lambda(\bk)$
appear in a simple BCS-like model in which the pairing interaction
is local in real space:
\begin{equation}
\label{Hint model}
    H_{int}=-\frac{V}{2}\sum_{\alpha\beta=\uparrow,\downarrow}
    \int d^3\br\,\psi_\alpha^\dagger(\br)\psi_\beta^\dagger(\br)
    \psi_\beta(\br)\psi_\alpha(\br),
\end{equation}
where $V>0$ is the coupling constant. Using the band
representation of the field operators, we obtain:
\begin{eqnarray}
\label{Hint model band gen}
    &&H_{int}=\frac{1}{2{\cal V}}\sum_{\bk\bk'\bq}\sum_{\lambda_{1,2,3,4}}
    V_{\lambda_1\lambda_2\lambda_3\lambda_4}(\bk,\bk')\nonumber\\
    &&\quad\times c^\dagger_{\bk+\bq/2,\lambda_1}
    c^\dagger_{-\bk+\bq/2,\lambda_2}c_{-\bk'+\bq/2,\lambda_3}c_{\bk'+\bq/2,\lambda_4},
    \quad
\end{eqnarray}
where
\begin{eqnarray*}
    V_{\lambda_1\lambda_2\lambda_3\lambda_4}(\bk,\bk')=-V
    \sum_{\alpha\beta}
    u^*_{\alpha\lambda_1}(\bk)u^*_{\beta\lambda_2}(-\bk)\\
    \times u_{\beta\lambda_3}(-\bk')u_{\alpha\lambda_4}(\bk').
\end{eqnarray*}
Here we replaced $u_{\alpha\lambda}(\pm\bk+\bq/2)$ by
$u_{\alpha\lambda}(\pm\bk)$, neglecting the corrections of the
order of $O(q/k_F)$. The expression (\ref{Hint model band gen})
contains both intra- and interband pairing terms. For the reasons
explained above, we neglect the latter and set
$\lambda_1=\lambda_2=\lambda$ and $\lambda_3=\lambda_4=\lambda'$,
which reduces the Hamiltonian to the form (\ref{H int}). Using the
identities
\begin{equation}
\label{k minus k}
    u_{\alpha\lambda}(-\bk)=t^*_\lambda(\bk)\sum_\beta(i\sigma_2)_{\alpha\beta}
    u^*_{\beta\lambda}(\bk),
\end{equation}
which follow from the definition (\ref{t_def}) of the phase factor
$t_\lambda(\bk)$, we obtain:
\begin{eqnarray*}
    V_{\lambda\lambda'}(\bk,\bk')&=&-Vt_\lambda(\bk)t_{\lambda'}^*(\bk')
    |w_{\lambda\lambda'}(\bk,\bk')|^2\nonumber\\
    &=&-Vt_\lambda(\bk)t^*_{\lambda'}(\bk')\frac{1+\lambda\lambda'\hat\bgam(\bk)\hat\bgam(\bk')}{2},
\end{eqnarray*}
see Eq. (\ref{w identity}). The terms in $\hat V$ containing the
$\hat\bgam$'s are even in both $\bk$ and $\bk'$. Using the
anti-commutation of the fermionic operators one can easily show
that these terms drop out of the Hamiltonian. Finally,
\begin{eqnarray}
\label{H int model}
    &&H_{int}=-\frac{V}{4}\frac{1}{{\cal V}}\sum\limits_{\bk\bk'\bq}\sum_{\lambda\lambda'}
    t_\lambda(\bk)t_{\lambda'}^*(\bk')\nonumber\\
    &&\qquad\times c^\dagger_{\bk+\bq/2,\lambda}c^\dagger_{-\bk+\bq/2,\lambda}
    c_{-\bk'+\bq/2,\lambda'}c_{\bk'+\bq/2,\lambda'}.
\end{eqnarray}
Thus the band representation of the pairing potential in the model
(\ref{Hint model}) necessarily contains the phase factors
$t_\lambda(\bk)$. The gap symmetry corresponds to the unity
representation with $\phi_\lambda(\bk)=1$, and all coupling
constants take the same value: $V_{\lambda\lambda'}=V/2$.

\subsection{Derivation of the free energy functional}
\label{sec: GL derivation}

The superconducting order parameter in the static case can be
represented as
\begin{equation}
\label{OP definition}
    \Delta_\lambda(\bk,\bq)=\eta_\lambda(\bq)t_\lambda(\bk)\phi_\lambda(\bk),
\end{equation}
where the bosonic fields $\eta_\lambda$ play the role of the order
parameter components. The free energy ${\cal F}$ (more precisely,
the difference between the free energies of the superconducting
and the normal states at the same temperature and field) is a
functional of $\eta_\lambda$. In the vicinity of the transition
temperature at arbitrary field the order parameter is small, and
one can keep in the free energy expansion only the terms quadratic
in $\eta_\lambda$. Following the procedure outlined, e.g. in Ref.
\onlinecite{Sam04}, we obtain the impurity-averaged free energy in
the form ${\cal F}={\cal F}_1+{\cal F}_2$, where
\begin{equation}
\label{F1 def}
    {\cal F}_1=\frac{1}{2}\sum\limits_{\lambda\lambda'}
    \frac{1}{{\cal V}}\sum_{\bq}\eta^*_\lambda(\bq)V^{-1}_{\lambda\lambda'}
    \eta_{\lambda'}(\bq),
\end{equation}
and
\begin{eqnarray}
\label{F2 def}
    {\cal F}_2&=&-\frac{1}{2}\sum_{\lambda\lambda'}T\sum_n
    \frac{1}{{\cal V}^2}\sum_{\bk\bk'\bq}\Delta^*_\lambda(\bk,\bq)
    \Delta_{\lambda'}(\bk',\bq)\nonumber\\
    &&\times\Bigl\langle\bar G_{\lambda\lambda'}\Bigl(\bk+\frac{\bq}{2},\bk'+\frac{\bq}{2};\omega_n\Bigr)
    \nonumber\\
    &&\times\bar G_{\lambda\lambda'}\Bigl(-\bk+\frac{\bq}{2},-\bk'+\frac{\bq}{2};-\omega_n\Bigr)
    \Bigr\rangle_{imp}.
\end{eqnarray}
This expression is applicable for any pairing symmetry described
by a one-dimensional representation of $\mathbb{G}$, and for any
band structure. There are corrections to ${\cal F}_2$ of the order
of $(T_c/\epsilon_F)^2$, related to the orbital magnetism of the
Cooper pairs,\cite{Book,Sam04} which we neglect. The GL gradient
expansion in all orders can be obtained from the Taylor expansion
of Eqs. (\ref{F1 def}) and (\ref{F2 def}) in powers of $\bq$, by
making the replacement
\begin{equation}
\label{q to D}
    \bq\to\bm{D}=-i\nabla_{\br}+\frac{2e}{\hbar c}\bm{A}(\br)
\end{equation}
in the final expressions.

\begin{figure}
    \includegraphics[width=8cm]{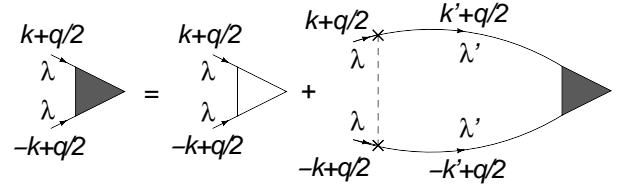}
    \caption{The diagrammatic representation of the gap equation.
    The triangle denotes the gap function $\Delta_\lambda$, the
    filled triangle -- the impurity-renormalized gap function
    $D_\lambda$, the dashed line corresponds to
    $W_{\lambda\lambda'}(\bk,\bk')$,
    and the solid lines are the average Green's functions of electrons.}
    \label{fig: gap eq}
\end{figure}

The disorder averaging in Eq. (\ref{F2 def}) involves summing the
impurity ladder diagrams, in which the average Green's function is
given by Eqs. (\ref{bar G average gen}), (\ref{bar G average}). In
the thermodynamic limit ${\cal V}\to\infty$, the momentum sums are
replaced by integrals, and we obtain:
\begin{eqnarray}
\label{F2 gen D}
    {\cal F}_2=-\frac{1}{2}\sum_{\lambda}\int\frac{d^3\bq}{(2\pi)^3}T\sum_n
    \int\frac{d^3\bk}{(2\pi)^3}\Delta^*_\lambda(\bk,\bq)\nonumber\\
    \times \bar G_{\lambda}\Bigl(\bk+\frac{\bq}{2},\omega_n\Bigr)
    \bar G_{\lambda}\Bigl(-\bk+\frac{\bq}{2},-\omega_n\Bigr)\nonumber\\
    \times D_\lambda(\bk,\bq,\omega_n),
\end{eqnarray}
where $D_\lambda$ are the impurity-renormalized gap functions,
which satisfy the following integral equations:
\begin{eqnarray}
\label{Delta eq gen}
    D_\lambda(\bk,\bq,\omega_n)=\Delta_\lambda(\bk,\bq)+
    \sum_{\lambda'}\int\frac{d^3\bk'}{(2\pi)^3}W_{\lambda\lambda'}(\bk,\bk')\nonumber\\
    \times\bar G_{\lambda'}\Bigl(\bk'+\frac{\bq}{2},\omega_n\Bigr)
    \bar G_{\lambda'}\Bigl(-\bk'+\frac{\bq}{2},-\omega_n\Bigr)\nonumber\\
    \times D_{\lambda'}(\bk',\bq,\omega_n),
\end{eqnarray}
with
\begin{eqnarray}
\label{W def}
    W_{\lambda\lambda'}(\bk,\bk')=\frac{1}{2\pi N_F\tau}
    w_{\lambda\lambda'}(\bk,\bk')w_{\lambda\lambda'}(-\bk,-\bk')\nonumber\\
    =\frac{1}{2\pi N_F\tau}t_\lambda(\bk)t^*_{\lambda'}(\bk')
    w_{\lambda\lambda'}(\bk,\bk')w_{\lambda'\lambda}(\bk',\bk)\nonumber\\
    =\frac{1}{2\pi N_F\tau}
    t_\lambda(\bk)t^*_{\lambda'}(\bk')\frac{1+\lambda\lambda'\hat\bgam(\bk)\hat\bgam(\bk')}{2}
\end{eqnarray}
corresponding to the impurity line, see Fig. \ref{fig: gap eq}. To
obtain the last expression we neglected the difference between
$\hat w(\bk+\bq/2,\bk'+\bq/2)$ and $\hat w(\bk,\bk')$ (this is
legitimate because the order parameter varies on the scale $q\ll
k_F^{-1}$, and $\hat w$ is a smooth function of $\bk,\bk'$ near
the Fermi surface), and also used the identities (\ref{k minus k})
and the fact that $\hat w^\dagger(\bk',\bk)=\hat w(\bk,\bk')$.

We seek solution of Eq. (\ref{Delta eq gen}) in the form
\begin{equation}
    D_\lambda(\bk,\bq,\omega_n)=t_\lambda(\bk)\bar\Delta_\lambda(\bk,\bq,\omega_n).
\end{equation}
Due to the momentum cutoff built in the basis functions,
$\bar\Delta_\lambda$ is restricted to the vicinity of the
$\lambda$th Fermi surface. Since the gap functions weakly depend
on energy in the vicinity of the Fermi surface, one can integrate
the products of two Green's functions with respect to
$\xi_\lambda=\xi_\lambda(\bk)$:
\begin{eqnarray*}
    N_\lambda\int d\xi_\lambda\bar G_{\lambda}\Bigl(\bk+\frac{\bq}{2},\omega_n\Bigr)
    \bar G_{\lambda}\Bigl(-\bk+\frac{\bq}{2},-\omega_n\Bigr)\\
    =\pi N_FL_{\lambda}(\bk,\bq,\omega_n),
\end{eqnarray*}
where
\begin{equation}
\label{L def}
    L_{\lambda}(\bk,\bq,\omega_n)=\frac{\rho_\lambda}{|\omega_n|
    +1/2\tau+i\Omega_{\lambda}(\bk,\bq)\sign\omega_n}
\end{equation}
depends on the direction of $\bk$. The notations are as follows:
\begin{equation}
\label{rho def}
    \rho_\lambda=\frac{N_\lambda}{N_F}
\end{equation}
is the fractional density of electronic states in the $\lambda$th
band ($0\leq\rho_\lambda\leq 2$, $\rho_++\rho_-=2$),
\begin{equation}
\label{Omega}
    \Omega_{\lambda}(\bk,\bq)=\frac{\hbar\bv_\lambda(\bk)\bq}{2}-\bm{m}_\lambda(\bk)\bB,
\end{equation}
and
$\bv_\lambda(\bk)=\hbar^{-1}\partial\xi_\lambda(\bk)/\partial\bk$
is the Fermi velocity. In this way we obtain the following
equation for $\bar\Delta_\lambda$ as a function of the direction
of $\bk$, at given $\bq$ and $\omega_n$:
\begin{eqnarray}
\label{barDelta eq}
    &&\bar\Delta_\lambda(\bk,\bq,\omega_n)=\eta_\lambda(\bq)\phi_\lambda(\bk)\nonumber\\
    &&\quad+\frac{1}{2\tau}\sum_{\lambda'}\biggl\langle
    \frac{1+\lambda\lambda'\hat\bgam(\bk)\hat\bgam(\bk')}{2}
    L_{\lambda'}(\bk',\bq,\omega_n)\nonumber\\
    &&\quad\times\bar\Delta_{\lambda'}(\bk',\bq,\omega_n)
    \biggr\rangle_{\lambda'}.
\end{eqnarray}
The phase factors $t_\lambda(\bk)$ drop out of the expression
(\ref{F2 gen D}) for ${\cal F}_2$, which takes the following form:
\begin{eqnarray}
\label{F 2 final}
    &&{\cal F}_2=-\frac{\pi N_F}{2}\int\frac{d^3\bq}{(2\pi)^3}\sum_{\lambda}
    \eta^*_\lambda(\bq)\nonumber\\
    &&\quad\times T\sum_n\Bigl\langle
    \phi_\lambda^*(\bk)L_{\lambda}(\bk,\bq,\omega_n)
    \bar\Delta_{\lambda}(\bk,\bq,\omega_n)\Bigr\rangle_{\lambda}.
\end{eqnarray}

We note that only the diagrams containing $\bar G_+\bar G_+$ and
$\bar G_-\bar G_-$ have been included in the impurity ladder in
Eq. (\ref{Delta eq gen}). The contribution from the diagrams with
$\bar G_+\bar G_-$ is smaller by a factor of
$\max(\varepsilon_c,\tau^{-1})/E_{SO}\ll 1$ and neglected.

The equations for $\bar\Delta_\lambda(\bk,\bq,\omega_n)$ can be
solved by transforming them into a system of linear algebraic
equations. We introduce
\begin{eqnarray}
\label{X def}
    &&X_a(\bq,\omega_n)\nonumber\\
    &&\quad=\sum_\lambda\left\langle \Lambda_{\lambda,a}(\bk)
    L_{\lambda}(\bk,\bq,\omega_n)\bar\Delta_{\lambda}(\bk,\bq,\omega_n)
    \right\rangle_{\lambda},
\end{eqnarray}
where $a=0,1,2,3$, and
$$
    \Lambda_{\lambda,a}(\bk)=\left\{\begin{array}{lll}
      1 & , & a=0 \\
      \lambda\hat\gamma_a(\bk) & , & a=1,2,3 \\
    \end{array}\right..
$$
Then Eqs. (\ref{barDelta eq}) can be written as
\begin{eqnarray}
\label{P eq 2}
    \bar\Delta_{\lambda}=\eta_\lambda\phi_\lambda+\frac{1}{4\tau}
    \sum_{a=0}^3\Lambda_{\lambda,a}X_a,
\end{eqnarray}
and the contribution (\ref{F 2 final}) to the free energy takes
the following form:
\begin{eqnarray}
\label{F2 Xa}
    {\cal F}_2&=&-\frac{N_F}{2}\int\frac{d^3\bq}{(2\pi)^3}\nonumber\\
    &&\times\biggl\{\sum_{\lambda}|\eta_\lambda(\bq)|^2\pi T\sum_n\left\langle
    |\phi_\lambda(\bk)|^2L_{\lambda}(\bk,\bq,\omega_n)\right\rangle_{\lambda}\nonumber\\
    &&+\frac{1}{4\tau}\sum_{\lambda}\eta^*_\lambda(\bq)\,\pi T\sum_n
    \sum_{a=0}^3X_a(\bq,\omega_n)\nonumber\\
    &&\times\left\langle\phi^*_\lambda(\bk)\Lambda_{\lambda,a}(\bk)L_{\lambda}(\bk,\bq,\omega_n)
    \right\rangle_{\lambda}\biggr\}.
\end{eqnarray}
From Eqs. (\ref{X def}) and (\ref{P eq 2}) we obtain a system of
four algebraic equations for $X_a(\bq,\omega_n)$:
\begin{equation}
\label{Xa eqs}
    \sum_{b=0}^3(\delta_{ab}-{\cal B}_{ab})X_b=Y_a,
\end{equation}
where
\begin{eqnarray*}
    &&{\cal B}_{ab}(\bq,\omega_n)=\frac{1}{4\tau}\sum_\lambda\left\langle
    \Lambda_{\lambda,a}(\bk)\Lambda_{\lambda,b}(\bk)L_{\lambda}(\bk,\bq,\omega_n)\right\rangle_{\lambda},\\
    &&Y_a(\bq,\omega_n)=\sum_\lambda \eta_\lambda(\bq)\left\langle\phi_\lambda(\bk)
    \Lambda_{\lambda,a}(\bk)L_{\lambda}(\bk,\bq,\omega_n)\right\rangle_{\lambda}.
\end{eqnarray*}

We see that, since $Y_a$ depend linearly on $\eta_\lambda$, so do
$X_a$, and the right-hand side of Eq. (\ref{F2 Xa}) is a bilinear
functional of the order parameter components. The total free
energy, which also includes the contribution (\ref{F1 def}), can
be written in the following form:
\begin{equation}
\label{F grad general}
    {\cal F}=\int\frac{d^3\bq}{(2\pi)^3}\sum\limits_{\lambda\lambda'}
    \eta^*_\lambda(\bq)f_{\lambda\lambda'}(\bq,\bB)\eta_{\lambda'}(\bq).
\end{equation}
The explicit expressions for the coefficients
$f_{\lambda\lambda'}$ can be derived in principle at arbitrary
$\bq$ and $\bB$, using the solution of the linear equations
(\ref{Xa eqs}). This procedure would allow one to calculate the
upper critical field $H_{c2}$ at any temperature. We focus on the
weak-field limit in the vicinity of the critical temperature, in
which case Eqs. (\ref{Xa eqs}) can be solved using a gradient
expansion, see Sec. \ref{sec: gradient exp} below.

We note that the free energy of our system is formally equivalent
to that of a two-band singlet superconductor, in which the
impurity scattering amplitude, the pairing symmetry, and the
Zeeman coupling are all anisotropic. Significant simplifications
are achieved, for example, in the case when the SO band splitting
is so large that there is only one nondegenerate band (say, the
``+''-band) crossing the Fermi level. This is formally described
by setting $\rho_-=0$ and $V_{+-}=V_{--}=0$. We will refer to this
case as the single-band limit. On the other hand, in the model
(\ref{Hint model}) the matrix $\hat V$ is degenerate and the whole
formalism should be modified. One can show that in this case the
pairing interaction (\ref{H int model}) can be decoupled using
just one bosonic field $\eta(\bq)$, so that the gap functions are
given by $\Delta_\lambda(\bk,\bq)=\eta(\bq)t_\lambda(\bk)$, and
the GL theory has the same form as in the standard, i.e.
one-component isotropic singlet, case. The physical properties of
these special cases will be studied below by taking the
appropriate limits in the general two-band expressions.

\subsection{Gradient expansion}
\label{sec: gradient exp}

The gradient and field expansion of the free energy (\ref{F grad
general}) is obtained by expanding both $L_\lambda$ and $X_a$ in
Eq. (\ref{F2 Xa}) in powers of $\Omega_\lambda$. In order to solve
Eq. (\ref{Xa eqs}) it is convenient to represent it in the
symbolic form $(1-\hat{\cal B})\bm{X}=\bm{Y}$. The $4\times 4$
matrix $\hat{\cal B}$ can be written as $\hat{\cal B}=\hat{\cal
B}_0+\delta\hat{\cal B}$, where in the first term $\Omega_\lambda$
is set to zero, while the second term contains the
$\Omega_\lambda$-dependent corrections. Similarly,
$\bm{Y}=\bm{Y}_0+\delta\bm{Y}$ and $\bm{X}=\bm{X}_0+\delta\bm{X}$,
where $\bm{X}_0=(1-\hat{\cal B}_0)^{-1}\bm{Y}_0$. Then,
\begin{eqnarray*}
    \delta\bm{X}&=&(1-\hat{\cal B}_0-\delta\hat{\cal B})^{-1}
    (\delta\bm{Y}+\delta\hat{\cal B}\bm{X}_0)\\
    &=&(1-\hat{\cal B}_0)^{-1}(\delta\bm{Y}+\delta\hat{\cal
    B}\bm{X}_0)\\
    &&+(1-\hat{\cal B}_0)^{-1}\delta\hat{\cal B}(1-\hat{\cal B}_0)^{-1}
    (\delta\bm{Y}+\delta\hat{\cal B}\bm{X}_0)+...
\end{eqnarray*}
Inserting this in Eq. (\ref{F2 Xa}), we obtain after some lengthy
but straightforward algebra:
\begin{equation}
\label{f expansion}
    \hat f=\hat f_0+\hat f_2+...,
\end{equation}
where $\hat f_m$ denotes the terms of the order of
$\Omega_\lambda^m$. The terms with odd powers of $\Omega_\lambda$
are proportional to $\sign\omega_n$ and vanish after the summation
over the Matsubara frequencies.

The uniform and field-independent contribution is given by
\begin{eqnarray}
\label{f0}
    A_{\lambda\lambda'}\equiv f_{0,\lambda\lambda'}=
    \frac{N_F}{2}\Bigl[\frac{1}{N_F}V^{-1}_{\lambda\lambda'}
    -\rho_\lambda\delta_{\lambda\lambda'}S_{01}\nonumber\\
    -\frac{1}{4\tau}\rho_{\lambda}\rho_{\lambda'}
    \langle\phi_{\lambda}\rangle\langle\phi_{\lambda'}\rangle
    S_{11}\Bigr].
\end{eqnarray}
To make the notations compact, here and below we omit the
arguments of the basis functions and the subscripts $\lambda$ in
the Fermi-surface averages, and also assume that the basis
functions are real. The Matsubara sums $S_{kl}$ are defined as
follows:
\begin{equation}
\label{Skl def}
    S_{kl}=\pi
    T\sum_n\frac{1}{|\omega_n|^k}\frac{1}{(|\omega_n|+1/2\tau)^l}.
\end{equation}
The logarithmically divergent sum in $S_{01}$ is cut off at
$n_c\simeq\varepsilon_c/2\pi T\gg 1$:
\begin{equation}
\label{S0}
    S_{01}=2\pi T\sum_{n=0}^{n_c}\frac{1}{\omega_n+1/2\tau}
    =\ln\frac{2\gamma\varepsilon_c}{\pi T}-\mathbb{F}(\tau T).
\end{equation}
Here $\ln\gamma\simeq 0.577$ is Euler's constant,
\begin{equation}
\label{mathbb F}
    \mathbb{F}(x)=\Psi\left(\frac{1}{2}+\frac{1}{4\pi x}\right)-\Psi\left(\frac{1}{2}\right),
\end{equation}
and $\Psi(x)$ is the digamma function.\cite{strong imp} One can
also check that $S_{11}=2\tau \mathbb{F}(\tau T)$. Therefore, we
have
\begin{eqnarray}
\label{A gen}
    A_{\lambda\lambda'}&=&\frac{N_F}{2}\Bigl[\frac{1}{N_F}V^{-1}_{\lambda\lambda'}
    -\rho_\lambda\delta_{\lambda\lambda'}
    \ln\frac{2\gamma\varepsilon_c}{\pi T}\nonumber\\
    &&+\Bigl(\rho_\lambda\delta_{\lambda\lambda'}-\frac{1}{2}\rho_{\lambda}\rho_{\lambda'}
    \langle\phi_{\lambda}\rangle\langle\phi_{\lambda'}\rangle
    \Bigr)\mathbb{F}(\tau T)\Bigr].
\end{eqnarray}

The second-order term in the expansion (\ref{f expansion}) has the
following form:
\begin{eqnarray}
\label{f2 gen}
    f_{2,\lambda\lambda'}(\bq,\bB)=\frac{N_F}{2}\Bigr[
    \rho_\lambda\delta_{\lambda\lambda'}\langle\phi^2_{\lambda}
    \Omega^2_{\lambda}\rangle S_{03}\nonumber\\
    +\frac{1}{4\tau}\rho_{\lambda}\rho_{\lambda'}
    \Upsilon_{\lambda\lambda'}(\bq,\bB)\Bigr],
\end{eqnarray}
where
\begin{eqnarray*}
    \Upsilon_{\lambda\lambda'}(\bq,\bB)=
    \left(\langle\phi_{\lambda}\rangle\langle\phi_{\lambda'}
    \Omega^2_{\lambda'}\rangle+\langle\phi_{\lambda'}\rangle\langle\phi_{\lambda}
    \Omega^2_{\lambda}\rangle\right)S_{13}\\
    +\frac{1}{4\tau}\langle\phi_{\lambda}\rangle\langle\phi_{\lambda'}\rangle
    \Bigl(\sum_{\lambda_1}\rho_{\lambda_1}\langle\Omega^2_{\lambda_1}\rangle\Bigr)
    S_{23}\\
    +\lambda\lambda'\sum_{i=x,y,z}\langle \hat\gamma_i\phi_{\lambda}\Omega_{\lambda}\rangle
    \langle \hat\gamma_i\phi_{\lambda'}\Omega_{\lambda'}\rangle \tilde
    S_{03,i}\\
    +\frac{1}{4\tau}\sum_{i=x,y,z}\Bigl(\sum_{\lambda_1}\lambda_1\rho_{\lambda_1}\langle\hat\gamma_i\Omega_{\lambda_1}
    \rangle\Bigr)\\
    \times\left(\lambda'\langle\phi_{\lambda}\rangle\langle\hat\gamma_i\phi_{\lambda'}\Omega_{\lambda'}\rangle
    +\lambda\langle\phi_{\lambda'}\rangle\langle\hat\gamma_i\phi_{\lambda}\Omega_{\lambda}\rangle\right)\tilde
    S_{13,i}\\
    +\frac{1}{(4\tau)^2}\langle\phi_{\lambda}\rangle\langle\phi_{\lambda'}\rangle
    \sum_{i=x,y,z}\Bigl(\sum_{\lambda_1}\lambda_1\rho_{\lambda_1}\langle\hat\gamma_i\Omega_{\lambda_1}
    \rangle\Bigr)^2\tilde S_{23,i},
\end{eqnarray*}
with
\begin{equation}
    \tilde S_{kl,i}=\pi T\sum_n\frac{1}{|\omega_n|^k}\frac{1}{(|\omega_n|+1/2\tau)^l}
    \frac{1}{|\omega_n|+s_i/2\tau},
\end{equation}
and
$s_i=1-\sum_\lambda\rho_\lambda\langle\hat\gamma_i^2(\bk)\rangle_\lambda/2$.

Along with the usual second-order gradient terms, the free energy
expansion also contains the terms linear in both $\bq$ and $\bB$,
which give rise to the magnetoelectric effect, see Refs.
\onlinecite{Lev85,Edel89,Edel95}, and also lead to the helical
superconducting phases considered in Refs.
\onlinecite{Agter03,Sam04,Kaur05}. It can be shown, however, that
these effects are small. Indeed, a typical contribution to $\hat
f_2$ is of the form $N_F\langle\Omega_\lambda^2\rangle/T_c^2$.
Minimizing it with respect to $\bq$, we find $q\sim\mu_BH/\hbar
v_F$, therefore the correction to the energy due to the Zeeman
term in Eq. (\ref{Omega}) is proportional to
$(\mu_BH/T_c)^2\sim(T_c/\epsilon_F)^2(m^*/m)^2(H/H_{c2})^2$, where
$m^*$ is the effective mass of quasiparticles and $H_{c2}$ is the
upper critical field at $T=0$. This ratio is typically very small
(unless the smallness of $T_c/\epsilon_F$ is compensated by a
large value of the effective mass), so it is legitimate to drop
the Zeeman terms and use
\begin{equation}
\label{Omega reduced}
    \Omega_{\lambda}(\bk,\bq)=\frac{\hbar\bv_\lambda(\bk)\bq}{2}.
\end{equation}
As a result, $\hat f_2$ becomes a second-degree polynomial in
$\bq$:
$f_{2,\lambda\lambda'}(\bq)=\sum_{ij}K_{\lambda\lambda',ij}q_iq_j$,
where the coefficients are found from Eq. (\ref{f2 gen}).

Finally, returning to real space and making the substitution
$\bq\to\bm{D}$, see Eq. (\ref{q to D}), we obtain the GL free
energy in the second order of the gradient expansion:
\begin{equation}
\label{GL energy reduced final}
    {\cal F}=\int d^3\br\sum\limits_{\lambda\lambda'}
    \eta^*_\lambda(\br)\Bigl[A_{\lambda\lambda'}+\sum_{ij}K_{\lambda\lambda',ij}
    D_iD_j\Bigr]\eta_{\lambda'}(\br),
\end{equation}
with $i,j=x,y,z$. The explicit expressions for the expansion
coefficients are determined by the pairing symmetry and given
below.

\emph{Unconventional pairing}: The order parameter transforms
according to a nontrivial one-dimensional representation of the
point group of the system, and
$\langle\phi_{\lambda}(\bk)\rangle_{\lambda}=0$. From Eq. (\ref{A
gen}) we obtain:
\begin{eqnarray}
\label{A uncon}
    A_{\lambda\lambda'}=\frac{N_F}{2}\Bigl[\frac{1}{N_F}V^{-1}_{\lambda\lambda'}
    -\rho_\lambda\delta_{\lambda\lambda'}\ln\frac{2\gamma\varepsilon_c}{\pi
    T}\nonumber\\
    +\rho_\lambda\delta_{\lambda\lambda'}\mathbb{F}(\tau T)\Bigr],
\end{eqnarray}
while many of the terms in Eq. (\ref{f2 gen}) vanish, giving
\begin{eqnarray}
\label{Kij uncon}
    &&K_{\lambda\lambda',ij}=\frac{\hbar^2N_F}{8}\Bigl[
    \rho_\lambda\delta_{\lambda\lambda'}\langle\phi^2_{\lambda}
    v_{\lambda,i}v_{\lambda,j}\rangle S_{03}\nonumber\\
    &&+\frac{1}{4\tau}\lambda\lambda'\rho_{\lambda}\rho_{\lambda'}\sum_{k=x,y,z}
    \langle\hat\gamma_k\phi_\lambda v_{\lambda,i}\rangle
    \langle\hat\gamma_k\phi_{\lambda'}v_{\lambda',j}\rangle\tilde
    S_{03,k}\Bigr].\qquad
\end{eqnarray}

\emph{Conventional pairing}:  The order parameter transforms
according to the trivial (unity) representation of the point
group. For simplicity, we consider only a fully isotropic pairing,
for which $\phi_{\lambda}(\bk)=1$, then
\begin{eqnarray}
\label{A con}
    A_{\lambda\lambda'}=\frac{N_F}{2}\Bigl[\frac{1}{N_F}V^{-1}_{\lambda\lambda'}
    -\rho_\lambda\delta_{\lambda\lambda'}\ln\frac{2\gamma\varepsilon_c}{\pi
    T}\nonumber\\
    +\frac{\lambda\lambda'}{2}\rho_+\rho_-\mathbb{F}(\tau
    T)\Bigr],
\end{eqnarray}
and
\begin{eqnarray}
\label{Kij con}
    &&K_{\lambda\lambda',ij}=\frac{\hbar^2N_F}{8}\biggl\{
    \rho_\lambda\delta_{\lambda\lambda'}\langle
    v_{\lambda,i}v_{\lambda,j}\rangle S_{03}\nonumber\\
    &&\quad+\frac{1}{4\tau}\rho_{\lambda}\rho_{\lambda'}
    \biggl[(\langle v_{\lambda,i}v_{\lambda,j}\rangle+\langle
    v_{\lambda',i}v_{\lambda',j}\rangle)S_{13}\nonumber\\
    &&\quad+\frac{1}{4\tau}(\rho_+\langle v_{+,i}v_{+,j}\rangle+\rho_-\langle v_{-,i}v_{-,j}\rangle)
    S_{23}\nonumber\\
    &&\quad+\lambda\lambda'\sum_{k=x,y,z}\langle\hat\gamma_k v_{\lambda,i}
    \rangle\langle\hat\gamma_k v_{\lambda',j}\rangle\tilde S_{03,k}\nonumber\\
    &&\quad+\frac{1}{4\tau}\sum_{k=x,y,z}(\rho_+\langle\hat\gamma_kv_{+,i}\rangle
    -\rho_-\langle\hat\gamma_kv_{-,i}\rangle)\nonumber\\
    &&\quad\times(\lambda\langle\hat\gamma_kv_{\lambda,j}\rangle+\lambda'\langle\hat\gamma_kv_{\lambda',j}\rangle)
    \tilde S_{13,k}\nonumber\\
    &&\quad+\frac{1}{(4\tau)^2}\sum_{k=x,y,z}(\rho_+\langle\hat\gamma_kv_{+,i}\rangle
    -\rho_-\langle\hat\gamma_kv_{-,i}\rangle)\nonumber\\
    &&\quad\times(\rho_+\langle\hat\gamma_kv_{+,j}\rangle-\rho_-\langle\hat\gamma_kv_{-,j}\rangle)
    \tilde S_{23,k}\biggr]\biggr\}.
\end{eqnarray}

In the next section we use the GL expansion (\ref{GL energy
reduced final}) to study the effects of impurities on the critical
temperature and the upper critical field.

\section{Calculation of observables}
\label{sec: observables}

\subsection{Critical temperature}
\label{sec: Tc}

At high temperatures the matrix $\hat A$ in Eq. (\ref{GL energy
reduced final}) is positive definite, therefore the minimum of the
free energy is achieved at $\eta_+=\eta_-=0$, which corresponds to
the normal state. The superconducting critical temperature $T_c$,
defined as the temperature below which one of eigenvalues of $\hat
A$ turns negative at zero field, is found from the equation
$\det\hat A=0$.

It is convenient to introduce the following notation:
\begin{equation}
\label{hat g def}
    g_{\lambda\lambda'}=N_FV_{\lambda\lambda'}\rho_{\lambda'}=V_{\lambda\lambda'}N_{\lambda'}.
\end{equation}
Since $\hat V$ is positive definite, we have $g_{++}>0$,
$g_{--}>0$, $\det\hat g>0$ (note that the matrix $\hat g$ is not
symmetric, in general). Using the expression (\ref{A gen}), the
equation for the critical temperature can be written as
$\det(\hat\tau_0+\hat g\hat M)=0$, where
\begin{eqnarray}
\label{M}
    M_{\lambda\lambda'}&=&-\delta_{\lambda\lambda'}
    \ln\frac{2\gamma\varepsilon_c}{\pi T_c}\nonumber\\
    &&+\left(\delta_{\lambda\lambda'}-\frac{\rho_{\lambda'}}{2}
    \langle\phi_{\lambda}\rangle\langle\phi_{\lambda'}\rangle\right)\mathbb{F}(\tau
    T_c).
\end{eqnarray}

At $\tau=\infty$ the second term in $M_{\lambda\lambda'}$
vanishes, and we obtain the critical temperature of the clean
superconductor:
\begin{equation}
\label{Tc0}
    T_{c0}=\frac{2\gamma\varepsilon_c}{\pi}e^{-1/g},
\end{equation}
where
\begin{equation}
\label{g def}
    g=\frac{g_{++}+g_{--}}{2}+\sqrt{\left(\frac{g_{++}-g_{--}}{2}\right)^2+g_{+-}g_{-+}}
\end{equation}
is the effective coupling constant.

We note that in the model (\ref{Hint model}), in which the pairing
is described by a single coupling constant
$V_{\lambda\lambda'}=V/2$, we have $g=N_FV$. Although the
expression for the critical temperature in this case has the usual
BCS form, the superconductivity is non-BCS, because the order
parameter resides in two nondegenerate bands (see the discussion
in the end of Sec. \ref{sec: GL derivation}), with the critical
temperature independent of the band splitting.

In the presence of impurities the cases of conventional and
unconventional pairing have to be considered separately.

\emph{Unconventional pairing:} $\langle\phi_{\lambda}\rangle=0$.
In this case, we obtain the following equation for $T_c$:
\begin{equation}
\label{Tc zero B uncon}
    \ln\frac{T_{c0}}{T_c}=\mathbb{F}(\tau T_c),
\end{equation}
where $\mathbb{F}(x)$ is defined by the expression (\ref{mathbb
F}). The reduction of the critical temperature is described by a
universal function, similar to the suppression of
superconductivity by paramagnetic impurities, see Ref.
\onlinecite{AG60}. In particular, at weak disorder, i.e. in the
limit $\tau T_{c0}\gg 1$, we have $\mathbb{F}(\tau
T_c)\simeq\pi/8\tau T_{c0}$, therefore
\begin{equation}
\label{Tc weak disorder uncon}
    T_c=T_{c0}-\frac{\pi}{8\tau}.
\end{equation}
Using the small-$x$ asymptotics
\begin{equation}
\label{F at large x}
    \mathbb{F}(x)=\ln\left(\frac{\gamma}{\pi
    x}\right)+\frac{2\pi^2}{3}x^2+O(x^3),
\end{equation}
we find from Eq. (\ref{Tc zero B uncon}) that the
superconductivity is completely suppressed at $\tau
T_{c0}=\gamma/\pi\simeq 0.567$.

\emph{Conventional pairing:} $\phi_{\lambda}=1$. Instead of Eq.
(\ref{Tc zero B uncon}) we obtain:
\begin{eqnarray}
\label{Tc zero B con}
    &&\ln\frac{T_{c0}}{T_c}\nonumber\\
    &&\quad=\frac{1+c_1\mathbb{F}(x)}{
    c_2+c_3\mathbb{F}(x)
    +\sqrt{c_4+c_5\mathbb{F}(x)
    +c_6\mathbb{F}^2(x)}}-\frac{1}{g},\qquad
\end{eqnarray}
where $x=\tau T_c$, and
\begin{eqnarray*}
    &&c_1=\frac{\rho_+(g_{--}-g_{+-})+\rho_-(g_{++}-g_{-+})}{2},\\
    &&c_2= \frac{g_{++}+g_{--}}{2},\\
    &&c_3=\frac{\det\hat g}{2},\\
    &&c_4=\left(\frac{g_{++}-g_{--}}{2}\right)^2+g_{+-}g_{-+},\\
    &&c_5=(c_2-c_1)\det\hat g,\\
    &&c_6=c_3^2.
\end{eqnarray*}

We see that the critical temperature depends on nonmagnetic
disorder, but in contrast to the unconventional case, the effect
is not described by a universal Abrikosov-Gor'kov function. At
weak disorder the suppression is linear in the scattering rate,
but with a non-universal slope:
\begin{equation}
\label{Tc weak disorder con}
    T_c=T_{c0}-\frac{1}{g}\left[c_1-\frac{1}{g}\left(c_3+
    \frac{c_5}{2\sqrt{c_4}}\right)\right]\frac{\pi}{8\tau}.
\end{equation}
In the opposite limit of strong impurity scattering, $\tau
T_{c0}\ll 1$ (but still $\tau\varepsilon_c\gg 1$, see Ref.
\onlinecite{strong imp}), we use $\mathbb{F}(x)=\ln(1/x)+O(1)\gg
1$ at $x\to 0$, to find that the critical temperature approaches
\begin{equation}
\label{Tc limit}
    T_c^*=T_{c0}\exp\left(\frac{1}{g}-\frac{c_1}{2c_3}\right),
\end{equation}
i.e. superconductivity is not completely destroyed by impurities.
The explanation is the same as in the conventional two-gap
superconductors, see e.g. Refs. \onlinecite{MP66,Kusa70,Gol97}:
Interband impurity scattering tends to reduce the difference
between the gap magnitudes in the two bands, which costs energy
and thus suppresses $T_c$, but only until both gaps become equal.
We have checked numerically that, for any positive-definite matrix
$\hat g$ and for any $\rho_\lambda$ satisfying the constraint
$\rho_++\rho_-=2$, both the coefficient in front of $\tau^{-1}$ in
Eq. (\ref{Tc weak disorder con}) and the exponent in Eq. (\ref{Tc
limit}) are negative, i.e. $T_c^*<T_{c0}$.

In the model (\ref{Hint model}), in which the pairing is isotropic
and the order parameter has only one component, there is an analog
of Anderson's theorem: The nonmagnetic disorder has no effect on
the critical temperature, since the right-hand side of Eq.
(\ref{Tc zero B con}) is identically zero. The same is also true
in the single-band limit, i.e. when the only nonzero constants are
$g_{++}=g$ and $\rho_+=2$.

\subsection{Upper critical field}
\label{sec: Hc2}

To illustrate the effect of disorder on the upper critical field,
we consider a tetragonal crystal, $\mathbb{G}=\mathbf{C}_{4v}$, in
the field $\bB=B\hat z$, in which case $H_{c2}(T)$ can be
calculated analytically, see also Ref. \onlinecite{Zhit04}. The GL
free energy (\ref{GL energy reduced final}) takes the form
\begin{eqnarray}
    {\cal F}&=&\int d^3\br\sum_{\lambda\lambda'}\eta^*_\lambda(\br)\Bigl\{
    A_{\lambda\lambda'}\nonumber\\
    &&+\left[K^\perp_{\lambda\lambda'}(D^2_x+D_y^2)
    +K^\parallel_{\lambda\lambda'}D_z^2\right]\Bigr\}\eta_{\lambda'}(\br),
\end{eqnarray}
where $\hat A,\hat K^\perp,\hat K^\parallel$ are real symmetric
matrices, see Eqs. (\ref{A con}) and (\ref{Kij con}) in the
conventional pairing case, and Eqs. (\ref{A uncon}) and (\ref{Kij
uncon}) in the unconventional pairing case.

In order to find the spectrum of the matrix differential operator
${\cal O}_{\lambda\lambda'}=K^\perp_{\lambda\lambda'}(D^2_x+D_y^2)
+K^\parallel_{\lambda\lambda'}D_z^2$, we introduce the operators
\begin{equation}
\label{a operators}
    a_\pm=\sqrt{\frac{\hbar c}{eB}}\frac{D_x\pm iD_y}{2},\quad
    a_3=\sqrt{\frac{\hbar c}{eB}}D_z.
\end{equation}
It is easy to check that $a_+=a_-^\dagger$ and $[a_-,a_+]=1$, and
therefore $a_\pm$ have the meaning of the raising and lowering
operators, while $a_3=a_3^\dagger$ commutes with both of them:
$[a_3,a_\pm]=0$. Representing ${\cal O}_{\lambda\lambda'}$ in
terms of the operators (\ref{a operators}), we have
\begin{equation}
\label{cal K a}
    {\cal O}_{\lambda\lambda'}=\frac{eB}{\hbar c}[K^\perp_{\lambda\lambda'}(4a_+a_-+2)
    +K^\parallel_{\lambda\lambda'}a_3^2].
\end{equation}
To calculate the eigenvalues explicitly, it is convenient to use
the basis of the Landau levels $|N,p\rangle$, which satisfy
\begin{eqnarray*}
    &&a_+|N,p\rangle=\sqrt{N+1}|N+1,p\rangle\\
    &&a_-|N,p\rangle=\sqrt{N}|N-1,p\rangle\\
    &&a_3|N,p\rangle=p|N,p\rangle,
\end{eqnarray*}
where $N=0,1,...$, and $p$ determines the modulation along the
$z$-axis: $p=k_z\sqrt{\hbar c/eB}$. Writing the order parameter as
a linear combination of the Landau levels,
$$
    \eta_\lambda(\br)=\sum_{N,p}\eta_{\lambda,N,p}\langle\br|N,p\rangle,
$$
we obtain:
\begin{equation}
    {\cal F}=\sum_{N,p}\sum_{\lambda\lambda'}
    {\cal L}_{\lambda\lambda'}(N,p)\eta^*_{\lambda,N,p}\eta_{\lambda',N,p},
\end{equation}
where
\begin{eqnarray}
    &&{\cal L}_{\lambda\lambda'}(N,p)\nonumber\\
    &&\qquad=A_{\lambda\lambda'}+\left[K^\perp_{\lambda\lambda'}(4N+2)
    +K^\parallel_{\lambda\lambda'}p^2\right]\frac{eB}{\hbar
    c}.\quad
\end{eqnarray}

The upper critical field $H_{c2}(T)$ can be found from the
equation $\det\hat{\cal L}=0$, after maximization with respect to
$N$ and $p$. We assume that the maximum is achieved at $N=p=0$ and
find the following expression for the slope of $H_{c2}(T)$ at
$B\to 0$:
\begin{eqnarray}
\label{Hc2 slope}
    R&\equiv&\left|\frac{dH_{c2}}{dT}\right|_{T=T_c}\\
    &=&\frac{\Phi_0}{2\pi}\left.\frac{A_{++}a_{--}+A_{--}a_{++}-2A_{+-}a_{+-}}{A_{++}K^\perp_{--}
    +A_{--}K^\perp_{++}-2A_{+-}K^\perp_{+-}}\right|_{T=T_c},\nonumber
\end{eqnarray}
where $\Phi_0=\pi\hbar c/e$ is the magnetic flux quantum, and
\begin{eqnarray*}
    a_{\lambda\lambda'}=\frac{N_F}{2T_c}\Bigl[\rho_\lambda\delta_{\lambda\lambda'}-
    \Bigl(\rho_\lambda\delta_{\lambda\lambda'}-\frac{1}{2}\rho_\lambda\rho_{\lambda'}
    \langle\phi_\lambda\rangle\langle\phi_{\lambda'}\rangle\Bigr)\\
    \times\frac{1}{4\pi\tau T_c}\Psi'\Bigl(\frac{1}{2}+\frac{1}{4\pi\tau T_c}\Bigr)\Bigr].
\end{eqnarray*}

While the expression (\ref{Hc2 slope}) is valid for arbitrary
disorder strength, we are especially interested in the limit of
weak disorder. At $\tau T_{c0}\gg 1$ the critical temperature can
be represented as $T_c=T_{c0}[1-b(\pi/8\tau T_{c0})]$, where the
dimensionless coefficients $b$ takes different values for
conventional and unconventional pairing, see Eqs. (\ref{Tc weak
disorder con}) and (\ref{Tc weak disorder uncon}), respectively.
The matrices $\hat A$, $\hat K^\perp$, $\hat a$ can also be
expanded in powers of $\tau^{-1}$:
\begin{equation}
\label{AKa}
    \hat A=\hat A_0+\delta\hat A,\
    \hat K^\perp=\hat K^\perp_0+\delta\hat K^\perp,\
    \hat a=\hat a_0+\delta\hat a,
\end{equation}
where
\begin{eqnarray*}
\label{matrices clean}
    &&A_{0,\lambda\lambda'}=\frac{N_F}{2}\Bigl(\frac{1}{N_F}V^{-1}_{\lambda\lambda'}
    -\rho_\lambda\delta_{\lambda\lambda'}\frac{1}{g}\Bigr),\\
    &&K^\perp_{0,\lambda\lambda'}=\delta_{\lambda\lambda'}\frac{7\zeta(3)\hbar^2N_F}{32\pi^2T_{c0}^2}
    \rho_\lambda\langle\phi_\lambda^2v_{\lambda,x}^2\rangle,\\
    &&a_{0,\lambda\lambda'}=\frac{N_F}{2T_{c0}}\rho_\lambda\delta_{\lambda\lambda'},
\end{eqnarray*}
$\zeta(z)$ is the Riemann zeta-function, and
\begin{eqnarray*}
    &&\delta A_{\lambda\lambda'}\nonumber\\
    &&\quad=\frac{N_F}{2}\left[(1-b)\rho_\lambda\delta_{\lambda\lambda'}
    -\frac{1}{2}\rho_{\lambda}\rho_{\lambda'}
    \langle\phi_{\lambda}\rangle\langle\phi_{\lambda'}\rangle\right]\frac{\pi}{8\tau T_{c0}},\quad\\
    &&\delta K^\perp_{\lambda\lambda'}\\
    &&\quad=\frac{\hbar^2N_F}{96T_{c0}^2}
    \biggl[\left(-3+\frac{42\zeta(3)b}{\pi^2}\right)\rho_\lambda\delta_{\lambda\lambda'}
    \langle\phi_\lambda^2v_{\lambda,x}^2\rangle\\
    &&\qquad+\frac{1}{2}\rho_{\lambda}\rho_{\lambda'}
    \Bigl(\langle\phi_{\lambda}\rangle\langle\phi_{\lambda'}v_{\lambda',x}^2\rangle+
    \langle\phi_{\lambda'}\rangle\langle\phi_{\lambda}v_{\lambda,x}^2\rangle\Bigr)\\
    &&\qquad+\frac{\lambda\lambda'}{2}\rho_{\lambda}\rho_{\lambda'}\sum_{i=x,y,z}
    \langle\hat\gamma_i\phi_\lambda v_{\lambda,x}\rangle
    \langle\hat\gamma_i\phi_{\lambda'}v_{\lambda',x}\rangle\biggr]\frac{\pi}{8\tau T_{c0}},\\
    &&\delta a_{\lambda\lambda'}=-\frac{1}{T_{c0}}\delta
    A_{\lambda\lambda'}
\end{eqnarray*}
[to get this we used $\zeta(4)=\pi^4/90$].

In the clean case, $A_{0,\lambda\lambda'}$ can be expressed in
terms of the matrix elements of $\hat g$, see Eq. (\ref{hat g
def}), which yields
\begin{equation}
\label{Hc2 slope clean}
    R_0=\frac{8\pi\Phi_0T_{c0}}{7\zeta(3)\hbar^2}\frac{g_{++}+g_{--}-2g^{-1}\det\hat g}{\sum_\lambda
    (g_{\lambda\lambda}-g^{-1}\det\hat
    g)\langle\phi_\lambda^2v_{\lambda,x}^2\rangle},
\end{equation}
with $g$ defined by Eq. (\ref{g def}). In the presence of
impurities, we substitute the expansions (\ref{AKa}) in Eq.
(\ref{Hc2 slope}) and obtain the correction to the upper critical
field slope:
\begin{eqnarray}
\label{delta R}
    \frac{\delta R}{R_0}&=&\biggl[\frac{{\cal D}_1}{g_{++}+g_{--}-2g^{-1}\det\hat g}
    \nonumber\\
    &&-\frac{{\cal D}_2}{\sum_\lambda(g_{\lambda\lambda}-g^{-1}\det\hat
    g)\langle\phi_\lambda^2v_{\lambda,x}^2\rangle}\biggr]\frac{\pi}{8\tau
    T_{c0}},\quad
\end{eqnarray}
where
\begin{widetext}
\begin{eqnarray*}
    {\cal D}_1&=&\frac{1}{2}(\rho_+g_{+-}+\rho_-g_{-+})\langle\phi_+\rangle\langle\phi_-\rangle
    -\sum_{\lambda=\pm}\left[g_{\lambda\lambda}-\left(1+\frac{1}{g}\right)\det\hat
    g\right]\left(1-b-\frac{\rho_\lambda}{2}\langle\phi_\lambda\rangle^2\right),\\
    {\cal D}_2&=&\frac{\pi^2}{21\zeta(3)}\biggl\{
    \sum_{\lambda=\pm}\left(g_{\lambda\lambda}-\frac{\det\hat g}{g}\right)\biggl[
    \left(\frac{42\zeta(3)b}{\pi^2}-3\right)
    \langle\phi_\lambda^2v_{\lambda,x}^2\rangle+\rho_\lambda\langle\phi_\lambda\rangle
    \langle\phi_\lambda v_{\lambda,x}^2\rangle+\frac{\rho_\lambda}{2}
    \sum_{i=x,y,z}\langle\hat\gamma_i\phi_\lambda v_{\lambda,x}\rangle^2\biggr]\nonumber\\
    &&+\frac{1}{2}(\rho_+g_{+-}+\rho_-g_{-+})\Bigl(\langle\phi_+\rangle
    \langle\phi_-^2v_{-,x}^2\rangle+\langle\phi_-\rangle
    \langle\phi_+^2v_{+,x}^2\rangle-\sum_{i=x,y,z}\langle\hat\gamma_i\phi_+v_{+,x}\rangle
    \langle\hat\gamma_i\phi_-v_{-,x}\rangle\Bigr)\biggr\}\nonumber\\
    &&+\det\hat g\left[\left(1-b-\frac{\rho_+}{2}\langle\phi_+\rangle^2\right)
    \langle\phi_-^2v_{-,x}^2\rangle+\left(1-b-\frac{\rho_-}{2}\langle\phi_-\rangle^2\right)
    \langle\phi_+^2v_{+,x}^2\rangle\right].
\end{eqnarray*}
\end{widetext}
It does not seem to be possible to draw any conclusions from Eq.
(\ref{delta R}) about the effect of impurities on the slope of
$H_{c2}$ in the general case, i.e. for arbitrary values of the
coupling constants  $g_{\lambda\lambda'}$ and the densities of
states $\rho_\lambda$. For this reason, we just look at two
limiting cases.

In the model (\ref{Hint model}), the expression (\ref{Hc2 slope
clean}) takes the form
\begin{equation}
\label{Hc2 slope model}
   R_0=\frac{16\pi\Phi_0T_{c0}}{7\zeta(3)\hbar^2}
   \frac{1}{\rho_+\langle v_{+,x}^2\rangle+\rho_-\langle
   v_{-,x}^2\rangle}.
\end{equation}
Neglecting the differences between the Fermi velocities and the
densities of states in the two bands (these differences are of the
order of $|\bgam|/\epsilon_F$) and assuming a spherical Fermi
surface, we obtain: $\rho_+\langle v_{+,x}^2\rangle+\rho_-\langle
v_{-,x}^2\rangle\to 2v_F^2/3$. In this way we recover Gor'kov's
expression for the slope of the upper critical field in a clean
isotropic superconductor.\cite{Gor59} In the same approximation,
Eq. (\ref{delta R}) becomes
\begin{equation}
\label{delta R model}
    \frac{\delta R}{R_0}=\frac{\pi^3}{168\zeta(3)}\frac{1}{\tau
    T_{c0}},
\end{equation}
which coincides with the one obtained in Ref. \onlinecite{Gor60}
for the impurity-induced enhancement of the upper critical field
in a conventional isotropic BCS superconductor.

In the single-band limit, setting
$g_{\lambda\lambda'}=g\delta_{\lambda,+}\delta_{\lambda',+}$,
$\rho_\lambda=2\delta_{\lambda,+}$, $\bv_+(\bk)=\bv(\bk)$,
$\phi_+(\bk)=\phi(\bk)$, we obtain from Eq. (\ref{Hc2 slope
clean}):
\begin{equation}
\label{Hc2 slope single-band}
   R_0=\frac{8\pi\Phi_0T_{c0}}{7\zeta(3)\hbar^2}
   \frac{1}{\langle\phi^2v_x^2\rangle},
\end{equation}
and from Eq. (\ref{delta R}):
\begin{equation}
\label{delta single-band}
    \frac{\delta R}{R_0}=C\frac{\pi^3}{168\zeta(3)}\frac{1}{\tau
    T_{c0}}.
\end{equation}
The coefficient $C$ is given by
\begin{equation}
\label{C con}
    C=C_{con}=1-\frac{\sum_i\langle\hat\gamma_iv_x\rangle^2}{\langle
    v_x^2\rangle}
\end{equation}
in the conventional pairing case, and
\begin{equation}
\label{C uncon}
    C=C_{uncon}=3-\frac{42\zeta(3)}{\pi^2}
    -\frac{\sum_i\langle\hat\gamma_i\phi v_x\rangle^2}{\langle
    \phi^2v_x^2\rangle}
\end{equation}
in the unconventional pairing case. Since $42\zeta(3)/\pi^2\simeq
5.115$, $C_{uncon}$ is always negative, i.e. the upper critical
field for unconventional pairing is reduced in the presence of
impurities. In the conventional case, the sign and the magnitude
of the correction  depend on the band structure. For example,
consider a cylindrical Fermi surface with
$\bv(\bk)=v_F(k_x,k_y,0)/k_\perp$, and
$\hat\bgam(\bk)=(k_y,-k_x,0)/k_\perp$ (the Rashba model), where
$k_\perp=\sqrt{k_x^2+k_y^2}$. Calculating the Fermi-surface
averages in Eq. (\ref{C con}), we find $C_{con}=1/2>0$, which
corresponds to the upper critical field enhancement by disorder.

\subsection{Application to CePt$_3$Si}
\label{sec: CePt3Si}

Application of our results to real noncentrosymmetric materials is
complicated by the lack of a definite information about the
superconducting gap symmetry and the distribution of the pairing
strength between the bands. One can only make progress by using
some simple models. For example, there are experimental
indications that most of the Fermi surface in CePt$_3$Si remains
normal.\cite{Bauer04} It is plausible that the superconductivity
resides in one of the two quasi-two-dimensional $\gamma$
bands.\cite{SZB04} Therefore we take into account only the
$\gamma$ bands and assume that, while the ``+''-band is
superconducting, the ``-''-band remains normal, i.e.
$V_{+-}=V_{--}=0$. In contrast to the purely single-band limit, in
the case under consideration $\rho_-\neq 0$.

As far as the pairing symmetry is concerned, there is strong
experimental evidence that the superconducting order parameter in
CePt$_3$Si has lines of gap
nodes.\cite{Yasuda04,Yogi04,Izawa05,Bonalde05,Tate04} The lines of
nodes are required by symmetry for all nontrivial one-dimensional
representations of $\mathbf{C}_{4v}$ ($A_2$, $B_1$, and $B_2$), so
that the superconductivity in CePt$_3$Si is most likely
unconventional. This can be verified using the measurements of the
dependence of $T_c$ on the impurity concentration: For all types
of unconventional pairing, the suppression of the critical
temperature is described by the universal Abrikosov-Gor'kov
function, see Eq. (\ref{Tc zero B uncon}). For the upper critical
field, one obtains from Eq. (\ref{delta R}):
\begin{equation}
\label{delta R CPS}
    \frac{\delta R}{R_0}=C\frac{\pi^3}{168\zeta(3)}\frac{1}{\tau
    T_{c0}},
\end{equation}
where
\begin{equation}
\label{C CPS}
    C=3-\frac{42\zeta(3)}{\pi^2}
    -\frac{\rho_+\sum_i\langle\hat\gamma_i\phi_+v_{+,x}\rangle^2}{2\langle
    \phi_+^2v_{+,x}^2\rangle}.
\end{equation}
One can see that for all types of unconventional pairing $C<0$,
i.e. the slope of $H_{c2}$ is reduced by disorder.

It should be mentioned that the lines of gap nodes can exist also
for conventional pairing ($A_1$ representation), in which case
they are purely accidental. While the accidental nodes would be
consistent with the power-law behavior of physical properties
observed experimentally, the impurity effect on $T_c$ in this case
is qualitatively different from the unconventional case. Indeed,
using Eq. (\ref{M}) we obtain the following equation for the
critical temperature:
\begin{equation}
\label{Tc eq CPS}
    \ln\frac{T_{c0}}{T_c}=\left(1-\frac{\rho_+}{2}\langle\phi_+\rangle^2\right)\mathbb{F}(\tau
    T_c),
\end{equation}
from which we have
\begin{equation}
\label{Tc CPS weak dis}
    T_c=T_{c0}\left(1-b\frac{\pi}{8\tau T_{c0}}\right),\quad
    b=1-\frac{\rho_+}{2}\langle\phi_+\rangle^2,
\end{equation}
at weak disorder ($\tau T_{c0}\gg 1$), and, using the asymptotical
expression (\ref{F at large x}),
\begin{equation}
\label{Tc CPS strong dis}
    T_c=T_{c0}\left(\frac{\pi\tau
    T_{c0}}{\gamma}\right)^\alpha,\quad\alpha=\frac{2}{\rho_+\langle\phi_+\rangle^2}-1,
\end{equation}
at strong disorder ($\tau T_{c0}\ll 1$). From the Schwarz
inequality $\langle\phi_+\rangle^2\leq\langle\phi_+^2\rangle=1$,
we find $b\geq\rho_-/2$ and $\alpha\geq\rho_-/\rho_+$. This means
that anisotropy of the conventional order parameter increases the
rate at which $T_c$ is suppressed by impurities. Unlike the
unconventional case, however, the superconductivity is never
completely destroyed, even at strong disorder.

\section{Conclusions}
\label{sec: Conclusion}

We have derived the Ginzburg-Landau free energy of a
noncentrosymmetric superconductor with isotropic nonmagnetic
impurities, using the microscopic model with a large spin-orbit
splitting of the electron bands. If the pairing corresponds to a
one-dimensional representation of the crystalline point group,
then the order parameter has two components, one for each band,
and the GL functional has the same form as for a two-band
superconductor (in the spin representation, the order parameter is
a mixture of spin-singlet and spin-triplet states, even without a
spin-triplet term in the pairing interaction, see Ref.
\onlinecite{Min04}).

We have also studied the impurity effect on the critical
temperature $T_c$ and the upper critical field $H_{c2}$. Although
$T_c$ is generally suppressed by impurities, this happens
differently for conventional and unconventional pairing: Any
deviation of $T_c(\tau)$ from the Abrikosov-Gor'kov curve, see Eq.
(\ref{Tc zero B uncon}), in particular an incomplete suppression
of superconductivity by strong disorder, is a signature of
conventional pairing symmetry. The impurity effect on the slope of
$H_{c2}$ turns out to be sensitive to the pairing symmetry and the
band structure, and therefore non-universal.

In the general case, i.e. for arbitrary values of the intraband
and interband coupling constants and the densities of states, the
microscopic expressions for the coefficients in the GL functional
are rather cumbersome and therefore of limited utility.
Considerable simplifications occur only in some cases, for
instance, in the BCS-like model (\ref{Hint model}), whose
properties resemble a conventional isotropic superconductor: The
order parameter has only one component, $T_c$ is not affected by
disorder, while $H_{c2}$ increases with disorder. On the other
hand, in CePt$_3$Si, where the pairing is likely unconventional,
with lines of the gap nodes, we obtain that both $T_c$ and
$H_{c2}$ are suppressed by disorder.

As far as experiment is concerned, the upper critical field
measurements in CePt$_3$Si, both in polycrystals\cite{Bauer04} and
in single crystals,\cite{Yasuda04} give the zero-temperature
values of $H_{c2}$ exceeding the Clogston-Chandrasekhar expression
for the paramagnetic limit, $H_P\simeq 1.24T_c/\mu_B$. This can be
understood as being due to the large residual spin paramagnetic
susceptibility, see Refs. \onlinecite{GR01,FAS04,Sam05}.
Surprisingly, the values of both the critical temperature and the
upper critical field in polycrystals are higher than in single
crystals. This is opposite to what has been observed in other
unconventional superconductors and also disagrees with our
theoretical predictions (assuming that the polycrystalline samples
are intrinsically ``more disordered'' than the single crystals).
To resolve this puzzle more work needs to be done on the purity
dependence of the properties of noncentrosymmetric
superconductors.

\section*{Acknowledgments}

The work of K.S. was financially supported by the Natural Sciences
and Engineering Research Council of Canada. K.S. is grateful to
the DRFMC/SPSMS for hospitality during his visit to CEA-Grenoble,
France, where this work was completed.


\begin{thebibliography}{99}
\bibitem{neglect orbital}
The orbital effect becomes important only at $T<\hbar\omega_c$
($\omega_c=eB/mc$ is the cyclotron frequency), which even in the
fields of the order of the upper critical field $H_{c2}$ means
$T<T_c^2/\epsilon_F$.
\bibitem{strong imp}
The expression (\ref{mathbb F}) for the impurity correction to
$S_{01}$ is valid if $\tau\varepsilon_c\gg 1$. In this case one
can extend the summation in $2\pi
T\sum_n[1/(\omega_n+1/2\tau)-1/\omega_n]$ to infinity and express
the result in terms of the digamma functions.
\end{thebibliography}
\end{document}